# Large-Signal Model of Graphene Field-Effect Transistors — Part II: Circuit Performance Benchmarking

Francisco Pasadas and David Jiménez

*Abstract*—This paper presents a circuit performance benchmarking using the large-signal model of graphene field effect transistor reported in Part I of this two-part paper. To test the model, it has been implemented in a circuit simulator. Specifically we have simulated a high-frequency performance amplifier, together with other circuits that take advantage of the ambipolarity of graphene, such as a frequency doubler, a radio-frequency subharmonic mixer and a multiplier phase detector. A variety of simulations comprising DC, transient dynamics, Bode diagram, S-parameters, and power spectrum have been compared with experimental data to assess the validity of the model.

## I. INTRODUCTION

THE growing interest in graphene electronics targeting radio-frequency (RF) and analog applications, results in a demand of graphene field effect transistor (GFET) compact models, which are needed to fill the gap between both device and circuit levels. Such models are embedded in a general purpose circuit simulator, which has to cope with DC, transient, and frequency response simulations of any graphene based circuit. Those circuit-compatible models could serve to different purposes depending on the Technology Readiness Level (TRL). For low TRL, compact models are useful in designing prototype devices/circuits, for device/circuit performance benchmarking against other technologies, and/or interpreting electrical measurements at the device/circuit level. If technology eventually became more mature (TRL higher), the compact models would be extremely useful to make the circuit design-fabrication cycle more efficient.

Manuscript received Month Day, Year; accepted Month Day, Year. Date of publication Month Day, Year; date of current version Month Day, Year. This work has received funding from the European Union's Horizon 2020 research and innovation programme under grant agreement No 696656 and from Ministerio de Economía y Competitividad under Grant TEC2012-31330 and Grant TEC2015-67462-C2-1-R.

The authors are with the Departament d'Enginyeria Electrònica, Escola d'Enginyeria, Universitat Autònoma de Barcelona, 08193 Bellaterra, Barcelona, Spain (e-mail: francisco.pasadas@uab.es, david.jimenez@uab.es).

Color versions of one or more of the figures in this paper are available online at http://ieeexplore.ieee.org.

Digital Object Identifier 10.1109/TED.2016.2563464

This paper is an extension to Part I [1] of a two-part paper. In Part I a large-signal model of GFETs has been developed. In this paper we have assessed our circuit-compatible model against experimental measurements. For such a purpose we have embedded the model in a Cadence Virtuoso Spectre Circuit Simulator [2], which is a widely used general purpose circuit simulator.

We have split the manuscript in two parts. For the first part we have assessed the DC and frequency response of a high-frequency voltage amplifier [3], which is a main building block of RF electronics. In the second part we have chosen exemplary circuits that take advantage of the graphene ambipolarity as the working principle. Specifically we have assessed the DC, transient dynamics, and frequency response of a high performance frequency doubler [4], a radio-frequency subharmonic mixer [5] and a multiplier phase detector [6].

## II. HIGH FREQUENCY PERFORMANCE OF GFETs

In this section, a high-frequency graphene voltage amplifier has been simulated and later compared with experimental results [3]. The GFET consists of a gate stack with an ultrathin high-κ dielectric (4 nm of $HfO_2$, equivalent oxide thickness EOT of 1.75 nm), which has been demonstrated to enhance current saturation [7]. The circuit under test is shown in Fig. 1, which is a common-source amplifier. The input parameters used for the GFET are described in Table I. The DC transfer characteristics and the GFET's transconductance are shown in Fig. 2a. Besides, the DC output characteristics at various gate-voltages are depicted in Fig. 2b.

TABLE I. INPUT PARAMETERS OF THE GFET USED TO SIMULATE THE VOLTAGE AMPLIFIER REPORTED IN [3].

| Input parameter | Value | Input parameter | Value |
|---|---|---|---|
| $T$ | 300 K | $L$ | 500 nm |
| $\mu$ | 4500 cm$^2$/Vs | $W$ | 30 μm |
| $V_{gs0}$ | 0.613 V | $L_t$ | 4 nm |
| $\varDelta$ | 0.095 eV | $\varepsilon_{top}$ | 12 |
| $\hbar\Omega$ | 0.12 eV | $R_s$, $R_d$ | 435 Ω·μm |
| $R_g$ | 14 Ω | | |

The meaning of the input parameters are explained in [1]. $T$ is temperature; $\mu$ represents the effective carrier mobility; $V_{gs0}$

is the top gate voltage offset; $\Delta$ is the inhomogeneity of the electrostatic potential due to electron-hole puddles; $\hbar\Omega$ is the effective energy at which a substrate optical phonon is emitted; $W$ and $L$ are the channel width and length, respectively; $L_t$ is the top oxide thickness; $\varepsilon_{top}$ refers the top oxide relative permittivity; and $R_s$, $R_d$ and $R_g$ refer to source, drain, and gate extrinsic resistances.

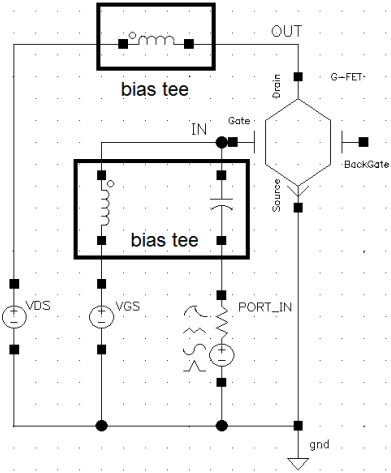

Fig. 1 Schematic circuit of the GFET based voltage amplifier. Bias tees are used for setting the DC bias point.

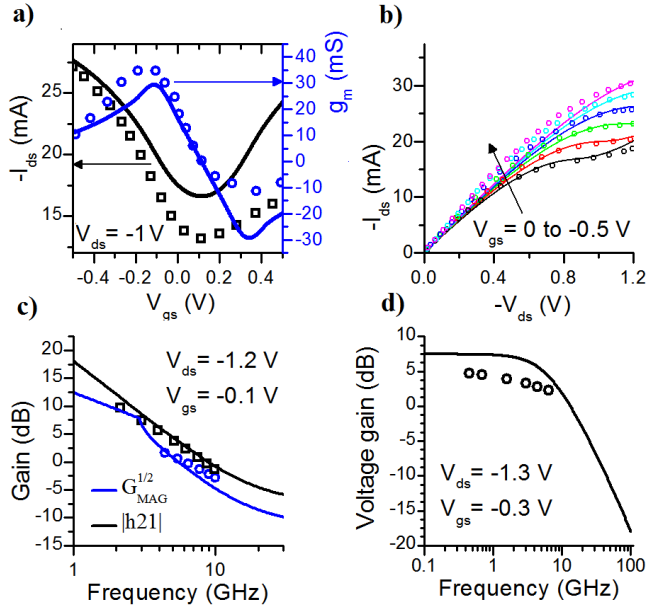

Fig. 2 (Color online) a) DC transfer characteristics and the extrinsic transconductance of the GFET based voltage amplifier. The device is biased at $V_{DS}$ = -1 V. b) DC output characteristics at various gate voltages. c) Power gain ($G_{MAG}^{1/2}$) and current gain ($|h_{21}|$) as a function of frequency. $f_{Tx}$ and $f_{max}$ are the frequency at which power gain and current gain becomes unity (0 dB), respectively. d) Frequency response of the amplifier's voltage gain when the input port level is -17 dBm.(Lines correspond to simulations and symbols to experimental data from [3])

Fig. 2c shows key RF characteristics of the GFET based voltage amplifier. For RF transistors, the cut-off frequency ($f_{Tx}$) and the maximum oscillation frequency ($f_{max}$) are the most widely used figures of merit (FoMs) to characterize the speed limit. The $f_{Tx}$ is defined as the frequency for which the magnitude of the small-signal current gain ($|h_{21}|$) of the transistor is reduced to unity. It is the highest possible frequency at which a FET is useful in RF applications. The simulation shows a $f_{Tx}$ = 8.7 GHz, in close agreement with the measured 8.2 GHz. On the other hand, the $f_{max}$ is defined as the highest possible frequency for which power gain ($G_{MAG}$), namely, the frequency where the magnitude of the power gain of the transistor is reduced to unity. The simulation shows a $f_{max}$ = 5.4 GHz that should be compared with the experimental 6.2 GHz. Finally, the voltage gain of the amplifier has been assessed (Fig. 2d). The simulation gives a DC voltage gain of ~ 7.4 dB, which is ~ $20\log(g_m g_{ds}^{-1})$, where $g_m$ (=$\partial I_{ds}/\partial V_{gs}$) is the transconductance and $g_{ds}$ (=$\partial I_{ds}/\partial V_{ds}$) is the output conductance, with a 3-dB bandwidth of 6.2 GHz.

### III. GRAPHENE-BASED AMBIPOLAR ELECTRONICS

Ambipolar electronics based on symmetric $I_{DS} - V_{GS}$ relation around the Dirac voltage ($V_{Dirac}$) has attracted lot of attention. The ability to control device polarity allows for i) redesign and simplification of conventional circuits such as frequency multipliers [4], [8]–[13], RF mixers [5], [10], [14]–[19], digital modulators [20]–[22], phase detectors [6] or active balun architectures [23]; and ii) opportunities for new functionalities in both analog/RF and digital domains. In this section we have benchmarked our model against exemplary ambipolar electronics' based circuits such as a high performance frequency doubler [4], a radio-frequency subharmonic mixer [5] and a multiplier phase detector [6].

#### A. Frequency doubler

The frequency doubler's working principle takes advantage of the quadratic behavior of the GFET transfer characteristics (TC), which can be written as

$$I_{DS} = a_0 + a_2 \left(V_{GS} - V_{Dirac}\right)^2 \quad (1)$$

where $a_0$ and $a_2$ are appropriate parameters describing the TC. When a small AC signal with an offset $V_{GS} = V_{Dirac}$, namely $V_{in} = V_{GS} + A\sin(\omega t)$, is input to the transistor's gate in the circuit of Fig. 3, the output voltage $V_{out} = V_{ds}$ results in:

$$V_{out} = V_{DD} - a_0 R_0 - \frac{1}{2} a_2 R_0 A^2 + \frac{1}{2} a_2 R_0 A^2 \cos(2\omega t) \quad (2)$$

where $A$ is the amplitude, $\omega = 2\pi f_{in}$ the angular frequency, and $R_0$ a load resistor connected to the drain. The output frequency is double because of the quadratic TC. If the TC was not perfectly parabolic and/or symmetric, which is the practical case, the output voltage would contain, in the former case, other even higher order harmonics and, in the latter case, other odd high order harmonics, resulting in harmonic distortion. Examples of frequency doublers can be found in [4], [8]–[12]. Moreover, with a properly adjusted threshold voltage separation of two graphene FETs connected in series, a graphene-based frequency tripler has been demonstrated [13].

Next, we proceed to apply the GFET model to the frequency doubler circuit shown in Fig. 3. The goal is to benchmark the model's outcome against the experimental data reported in [4]. The input parameters used for the GFET are shown in Table II. The DC transfer characteristics and the GFET transconductance, are shown in Fig. 4a, with a nearly symmetric shape respect to the Dirac voltage, $V_{Dirac} = -1.15$ V.

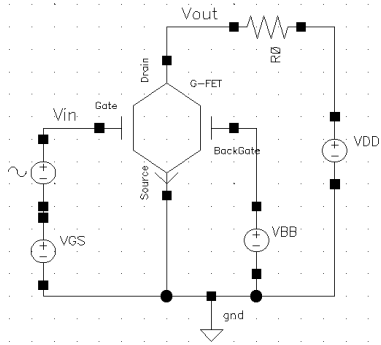

Fig. 3 Schematic circuit of the GFET based frequency doubler.

TABLE II. INPUT PARAMETERS OF THE GFET USED TO SIMULATE THE CIRCUIT REPORTED IN [4].

| Input parameter | Value | Input parameter | Value |
| --- | --- | --- | --- |
| $T$ | 300 K | $L$ | 500 nm |
| $\mu$ | 1300 cm$^2$/Vs | $W$ | 840 nm |
| $V_{gs0}$ | -1.062 V | $L_t$ | 5 nm |
| $V_{bs0}$ | 0 V | $L_b$ | 300 nm |
| $\Delta$ | 0.140 eV | $\varepsilon_{top}$ | 12 |
| $\hbar\Omega$ | 0.075 eV | $\varepsilon_{bottom}$ | 3.9 |
| $R_s, R_d$ | 1.1 kΩ·μm | $R_g$ | 20 Ω |

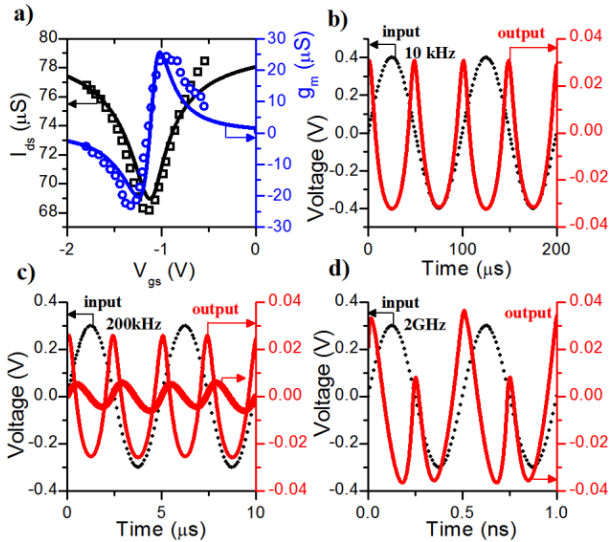

Fig. 4 (Color online) a) DC transfer characteristics and the extrinsic transconductance of the GFET based frequency doubler. The device is biased at $V_{DD}$ = 1 V, $V_{BB}$ = 40 V and $V_{GS}$ = -1.15 V. b) Input and output waveforms considering an input frequency of $f_{in}$ = 10 kHz and amplitude $A$ = 400 mV. c) Input and output waveforms considering an input frequency of $f_{in}$ = 200 kHz and amplitude $A$ = 300 mV. A thicker solid line shows the output waveform when a parasitic capacitance ($C_{pad}$ = 600 pF) is placed between the drain-source and the back-gate, taking into account the effect of the electrode pads. d) Input and output waveforms considering an input frequency of $f_{in}$ = 2 GHz and amplitude $A$ = 300 mV.

Using the GFET model we have analyzed the output waveform for different input frequencies, which are shown in Fig. 4b-d. For the lowest frequency, $f_{in}$ = 10 kHz, the output waveform consists of the doubled frequency with an amplitude ~ $A/10$, with a clear distortion coming from other higher order harmonics (see Fig. 4b). A Fourier transform of the waveform, shown in Fig. 5, reveals that 60% of the output RF power is concentrated at the doubled frequency of 20 kHz.

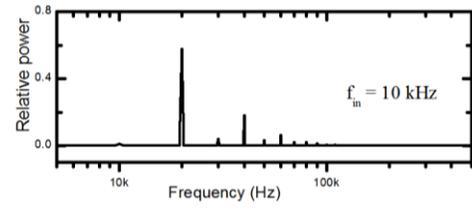

Fig. 5 Power spectrum obtained via Fourier transforming the output signal in Fig. 4b.

When the input signal is increased up to $f_{in}$ = 200 kHz and beyond a severe decay of the output signal amplitude was observed in the experiment, with a voltage gain ~ $A/100$ [4], likely because of the presence of a parasitic capacitance $C_{pad}$ = 600 pF between the GFET source-drain terminals and its back-gate, getting a similar output waveform as in the experiment for an input frequency of 200 kHz (see Fig. 4c). If the input frequency is further increased up to 2 GHz the output waveform, shown in Fig. 4d, displays the doubled frequency, although with a greater distortion because the group delay is not constant with the frequency according to Fig. 6, meaning that the phase is not linear with the frequency. To achieve high efficiency gigahertz frequency multipliers the parasitic capacitances must be diminished. Besides, these non-idealities must be incorporated to the device model to make realistic predictions on the performance of high frequency circuits.

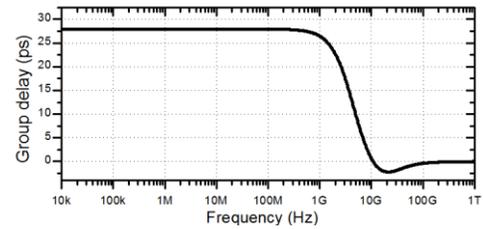

Fig. 6 Group delay vs. frequency for the GFET based frequency doubler.

### B. RF mixer

In telecommunications, a mixer is a nonlinear device that receives two different frequencies (the local oscillator LO signal at $f_{LO}$ and the radio-frequency RF signal at $f_{RF}$) at the input port and a mixture of several frequencies appears at the output, including both original input frequencies, the sum of the input frequencies, the difference between the input frequencies (the intermediate frequency IF signal at $f_{IF}$), and other intermodulations [24]. There are basically two operating principles for a FET mixer; either utilizing the change in transconductance, $g_m$, or channel conductance, $G_{ds}$ ($= I_{DS}/V_{DS}$), with the gate voltage. In both approaches a LO signal is applied to the gate to achieve a resulting time-varying, periodic quantity $g_m(t)$ or $G_{ds}(t)$. The former case is referred to as an active transconductance mixer, where the RF signal is applied to the gate, and the latter a resistive mixer, with the RF signal applied to the drain [18].

On the one hand, best possible performance from a transconductance mixer is realized by maximizing the variation in $g_m$, which is accomplished by biasing the FET in the saturation region. Examples of graphene-based transconductance mixers can be found in [10], [14]. However, as a consequence of the currently low transconductance in GFETs and the weak current saturation, the so far reported graphene-based transconductance mixers have shown poor performance. Instead, it does seem better to use the resistive mixing concept combined with the unique properties of graphene allowing for the design of subharmonic mixers with a single FET. The mixer operation is based on a sinusoidal LO signal also applied to the gate of the GFET, biased at the Dirac voltage. The idea is to make a frequency doubler operation with the LO signal as explained in subsection III.A, but keeping the drain unbiased. Thus the conductance variation as seen from the drain, $G_{ds}(t)$ would have a fundamental frequency component twice as $f_{LO}$. Therefore a subharmonic mixer only needs half the LO frequency compared to a fundamental mixer. This property is attractive particularly at high frequencies where there is a lack of compact sources providing sufficient power [25]. Moreover, subharmonic mixers suppress the LO noise [26], and the wide frequency gap between the RF and LO signals simplifies the LO and RF separation [27]. Examples of resistive mixers without subharmonic operation are reported in [15]–[17] and examples of resistive subharmonic mixers can be found in [5], [18], [19]. Besides, due to near symmetrical ambipolar conduction, graphene-based mixers can effectively suppress odd-order intermodulations, which are often present in conventional unipolar mixers and are harmful to circuit operations [28].

low-pass filter, both assumed with cutoff frequencies of 800 MHz and 30 MHz, respectively.

TABLE III. INPUT PARAMETERS OF THE GFET USED TO SIMULATE THE CIRCUIT REPORTED IN [5].

| Input parameter | Value | Input parameter | Value |
|---|---|---|---|
| $T$ | 300 K | $L$ | 1 µm |
| $\mu$ | 2200 cm$^2$/Vs | $W$ | 20 µm |
| $V_{gs0}$ | 1 V | $L_t$ | 25 nm |
| $V_{bs0}$ | 0 V | $L_b$ | 300 nm |
| $\Delta$ | 0.116 eV | $\varepsilon_{top}$ | 9 |
| $\hbar\Omega$ | 0.075 eV | $\varepsilon_{bottom}$ | 3.9 |
| $R_s$, $R_d$ | 560 Ω·µm | $R_g$ | 20 Ω |

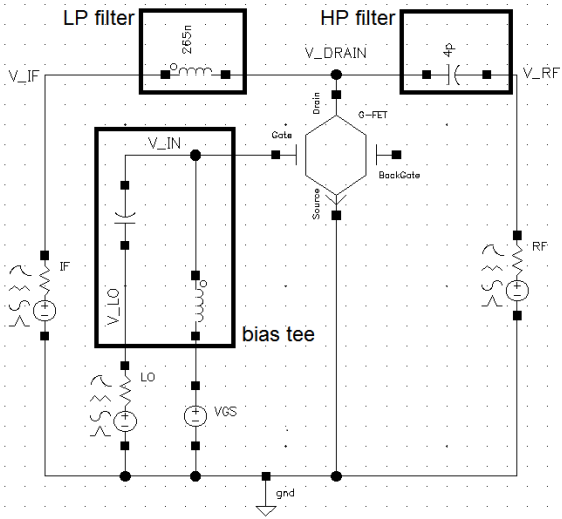

Fig. 7 Schematic circuit of the subharmonic resistive GFET mixer. A bias tee is used for setting the DC bias point. The characteristic impedance of the ports is $Z_0 = 50$ Ω.

In this subsection, we have applied the GFET model to the subharmonic resistive mixer circuit shown in Fig. 7. The goal is to benchmark the model's outcome against the experimental data reported in [5]. The input parameters used for the GFET are shown in Table III. The circuit under test only uses a transistor and no balun is required in that implementation, which makes the mixer more compact, as opposed to conventional subharmonic resistive FET mixers, which require two FETs in a parallel configuration, including a balun for feeding the two out-of-phase LO signals [29], [30]. In the subharmonic mixer, the RF signal is applied to the drain of the GFET through a high-pass filter and the IF is extracted with a

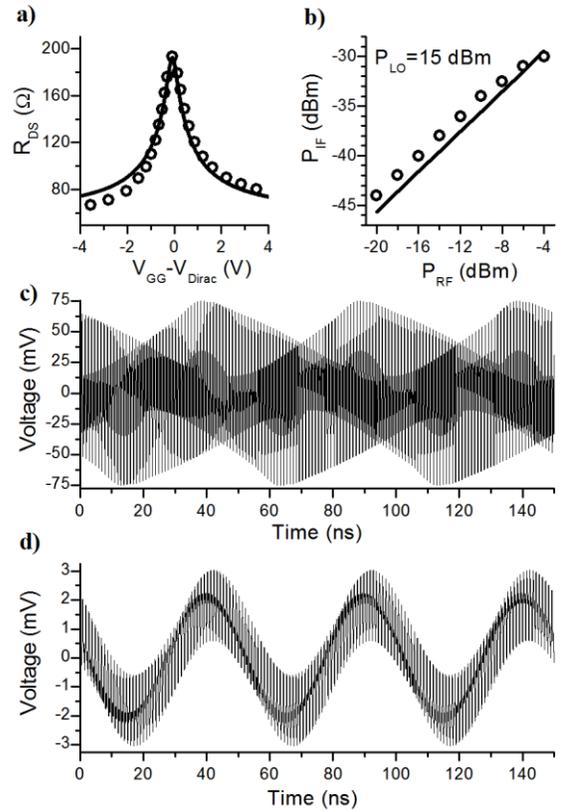

Fig. 8 a) Drain-to-source resistance $R_{DS} = 1/G_{DS}$ versus the gate voltage $V_{GS}$, with $R_{DS} = R_d + R_s + R_{ch}$, where $R_{ch}$ is the channel resistance and $R_d$ and $R_s$ are the extrinsic contact resistances at the drain and source sides. Solid lines correspond to simulations and the symbols to the experimental results in [5]. b) IF output power as a function of the RF input power. The device is biased at $V_{GS} = V_{Dirac}$ and $P_{LO} = 15$ dBm. c) Transient evolution of the signal collected at the drain at $V_{GS} = V_{Dirac}$. The following conditions have been assumed: $P_{LO} = 15$ dBm and $f_{LO} = 1.01$ GHz; $P_{RF} = -20$ dBm and $f_{RF} = 2$ GHz. d) Transient evolution of the IF signal collected at the IF port under the same conditions as in c). The separation between peaks is 50 ns, which corresponds to $f_{IF} = |f_{RF} - 2f_{LO}| = 20$ MHz.

The drain-to-source resistance $R_{DS} = 1/G_{DS}$ versus the gate voltage is shown in Fig. 8a. The device has been bias at $V_{GS} = V_{Dirac} = 1$ V through a bias tee. The RF signal has been introduced to RF port connected to the drain and the LO signal has been introduced to the LO port connected to the gate through the bias tee, where the IF signal is collected at the IF port, according to the schematics shown in Fig. 7. Fig. 8b depicts the mixer IF output power versus the RF input power, where a near constant conversion loss rate of ~ 25 dB has been obtained. The transient evolution of the signal collected at the drain is shown in Fig. 8c, as well as the signal collected at the IF port (Fig. 8d), which oscillates as expected at $f_{IF} = |f_{RF} - 2f_{LO}| = 20$ MHz. Finally, the spectrum of the signal collected at the drain is represented in Fig. 9, being the output power of ~ -49 dBm. Lower levels of odd harmonics are observed as well, which are attributed to the non-perfect symmetry of $R_{DS}$ versus $V_{GS}$.

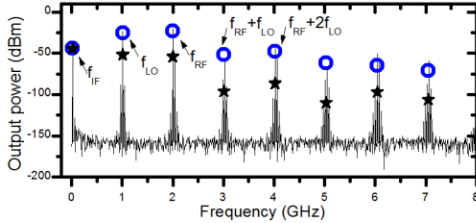

Fig. 9 (Color online) Spectrum (solid lines) of the signal collected at the drain ($P_{LO} = 15$ dBm and $f_{LO} = 1.01$ GHz; $P_{RF} = -20$ dBm and $f_{RF} = 2$ GHz). The bubbles correspond to the experimental results in [5]; and the stars correspond to the power peaks of the signal collected at the IF port.

*C. Multiplier phase detector*

The multiplier phase detector is a vital component of the phase-locked loop, which is one of the most important building blocks in modern analog, digital, and communication circuits [31].

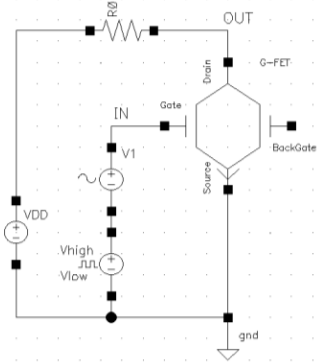

Fig. 10 Schematics of the multiplier phase detector based on a single graphene transistor and a load resistor.

Upon application of a sinusoidal wave $A_1\sin(\omega t+\theta_1)$ and a square wave $A_2\text{rect}(\omega t+\theta_2)$ to the input of a phase detector, the DC component of the output can be written as the product of the two input signals [6]:

$$A_d = A_1 A_2 \frac{2}{\pi} \sin(\theta_1 - \theta_2) \simeq K_d \theta_e \quad (3)$$

where $K_d$ is the gain of the detector and $\theta_e$ is the phase difference in radians between the input signals. Hence, the relation between the DC component and the phase difference can be utilized for phase detection. A multiplier is generally needed for this process, which complicates the circuit. However, taking advantage of the ambipolarity of a GFET, the simplified circuit structure shown in Fig. 10 is enough to perform the phase detection.

Next, we proceed to apply the GFET model to the phase detector circuit shown in Fig. 10 with the goal of benchmarking the model's outcome against the experimental data reported in [6]. The input parameters used for the GFET are shown in Table IV. The DC transfer characteristics and the GFET's transconductance at $V_{ds} = 0.1$ V are shown in Fig. 11a. The device shows a nearly symmetric characteristic around the Dirac voltage ($V_{Dirac} = 0.55$ V). Then, the GFET is biased at $V_{DD} = 1.8$ V through a series resistor $R_0 = 20$ kΩ, according to the schematics shown in Fig. 10. The back-gate has been assumed disconnected, as in [6]. A square-wave signal is used as the gate bias voltage, where the lower ($V_{low} = 0.36$ V) and ($V_{high} = 0.82$ V), satisfy $V_{low} < V_{Dirac}$ and $V_{high} > V_{Dirac}$. Both levels match with the two $g_m$ peaks so to get the maximum voltage gain. A sinusoidal-wave signal with 0.1 V of amplitude oscillates around the two levels of the square-wave signal. Both signals have 100 kHz of frequency, thus resulting in the following combined gate input signal:

$$v_{IN} = 0.1\sin(2\pi 10^5 t + \theta_1) + [0.46\text{rect}(2\pi 10^5 t + \theta_2) + 0.36] \quad (4)$$

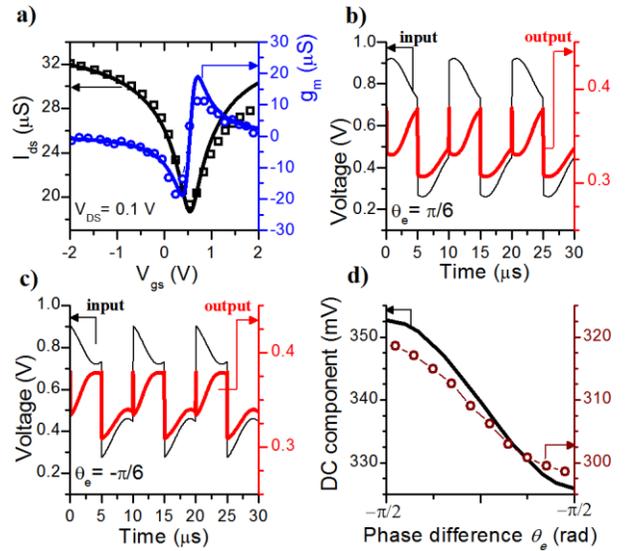

Fig. 11 (Color online) a) Experimental (symbols) and simulated (solid lines) DC transfer characteristics and extrinsic transconductance of the device described in Table IV. The bias has been set to $V_{DS} = 0.1$ V. b, c) Simulated input and output waveforms in the phase detector circuit shown in Fig. 10, biased at $V_{DD} = 1.8$ V, where a phase difference $\theta_e = \pi/6$ (b) and $\theta_e = -\pi/6$ (c) has been assumed. The transient responses are quite similar to the data reported in [6]. d) Experimental (symbols) and simulated (solid line) output DC component versus the phase difference $\theta_e$.

In Fig. 11b and Fig. 11c the transient response of the multiplier phase detector circuit have been depicted, assuming both $\theta_e = \pi/6$ and $\theta_e = -\pi/6$, respectively, which look very similar to the experimental results. The circuit corresponds to a common-source amplifier, therefore the voltage gain could be estimated as $A_v \approx -g_m(g_{ds}^{-1}\|R_0)$. It is approximately 0.1, which agrees with the reported value in [6]. Finally, in Fig. 11d, the output DC component is shown for different $\theta_e$. As the phase difference goes from $-\pi/2$ to $\pi/2$ rad, the DC component decreases from 353 to 326 mV, which corresponds to a detector gain of $K_d \approx -8.6$ mV/rad, which can be further improved by combining a reduction of the series resistance, increasing the gate efficiency (increase $g_m$), and pushing the transistor to saturation region (reducing $g_{ds}$).

TABLE IV. INPUT PARAMETERS OF THE GFET USED TO SIMULATE THE PHASE DETECTOR REPORTED IN [6]

| Input parameter | Value | Input parameter | Value |
|---|---|---|---|
| $T$ | 300 K | $L$ | 1.28 μm |
| $\mu$ | 2100 cm$^2$/Vs | $W$ | 2.98 μm |
| $V_{gs0}$ | 0.495 V | $L_t$ | 23 nm |
| $V_{bs0}$ | 0 V | $L_b$ | 300 nm |
| $\Delta$ | 0.074 eV | $\varepsilon_{top}$ | 9.35 |
| $\hbar\Omega$ | 0.075 eV | $\varepsilon_{bottom}$ | 3.9 |
| $R_s$, $R_d$ | 4.3 kΩ·μm | $R_g$ | 30 Ω |

## IV. CONCLUSIONS

In conclusion, we have presented a large-signal GFET model suitable for circuit design [1] (Verilog-A version available online at http://ieeexplore.ieee.org) and it has been benchmarked against high-performance and ambipolar electronics' circuits. The agreement between experiment and simulation is quite good in general, although fine adjustment would require further modeling of parasitic effects such as extrinsic capacitances and voltage-dependent contact resistances. The GFET model is compatible with conventional circuit simulators allowing for technology benchmarking, performance metrics prediction and design of circuits offering new functionalities. The intrinsic description of the device serves as a starting point toward a complete GFET model that could incorporate additional non-idealities.

## ACKNOWLEDGMENTS

This work has received funding from the European Union's Horizon 2020 research and innovation programme under grant agreement No 696656 and from Ministerio de Economía y Competitividad under Grant TEC2012-31330 and Grant TEC2015-67462-C2-1-R. We also acknowledge Xavier Cartoixà for the technical help.